# АНАЛИЗ СИММЕТРИЙ ГЕНЕТИЧЕСКОГО КОДА И СТЕПЕНЬ ДЕТЕРМИНАЦИИ КОДОНОВ

**Д. Р. Дуплий, С. А. Дуплий** *

*Харьковский национальный университет им. М. Н. Каразина*
*пл. Свободы, 4, Харьков 61077, Украина*

22 июня 2000 г.

Предлагается трехмерная модель генетического словаря в терминах введенной числовой характеристики нуклеотидов — степени детерминации, которая отражает абсолютную разность содержания пиримидиновых и пуриновых оснований в одной нити ДНК. В рамках этой модели прослеживаются выраженные симметрийные закономерности и групповые свойства, которые описаны. Предлагается использовать степень детерминации для анализа генетических текстов а также предсказания структур и значения различных функциональных участков ДНК.

*E-mail: Steven.A.Duplij@univer.kharkov.ua. Internet: http://gluon.physik.uni-kl.de/~duplij

Понимание современных концепций молекулярной генетики невозможно без дальнейшего всестороннего изучения аппарата экспрессии генов [1, 2]. Весьма актуальным в этом направлении являются алгебраические подходы к проблеме эволюции эукариотического генетического кода [3, 4], заключающиеся в применении теоретико-групповых методов [5], поиске симметрий [6, 7] и суперсимметрий [8, 9, 10], техники векторных пространств [11]. Однако внутренняя структура генетического кода и топология ДНК требуют дальнейшего исследования [12, 13, 14].

В работе предлагается трехмерная модель генетического словаря, на основании разной степени детерминации нуклеотидов $d$, которая описывается количественно. Внутри полученной модели прослеживаются выраженные симметрийные закономерности и групповые свойства. Введенная степень детерминации нуклеотида $d$ отражает абсолютную разность содержания пиримидиновых и пуриновых оснований в наборах нуклеотидов и находится с ней в периодической зависимости. Обсуждается возможность использования этой числовой характеристики для анализа и чтения генетических текстов, а также предсказание строения различных функциональных участков ДНК.

## 1 Вырожденность кода и специфичность нуклеотидов

Хорошо известно, что генетический код имеет триплетный характер с разной степенью специфичности оснований [15]. Так первые два основания кодона в большей степени детерминируют образование определенной аминокислоты, чем третье основание. Например, любая из аминокислот — глицин, валин, пролин, аланин и треонин — кодируется четырьмя кодонами, и в каждом случае эти четыре кодона различаются только нуклеотидами в третьей позиции. Другими словами 32 кодона,то есть половина всего количества, имеет полное вырождение по третьему основанию $z$. Аминокислота полностью задана первыми нуклеотидами $x$ и $y$ независимо от нуклеотида $z$. Для двукратно вырожденных кодонов характер аминокислоты определяется пуринами ($z = A, G$) или пиримидинами ($z = C, U$) находящимися в третьей позиции. Около двух третей общего количества оснований, присутствующих в ДНК имеют примерно постоянный характер у всех организмов — это основания, занимающие первое и второе положение в триплете, изменчивость состава ДНК определяется основаниями в третьем положении [16, 15].

Корреляция между количеством определенных аминокислот в белке и относительным содержанием гуанина и цитозина в соответствующей ДНК была найдена еще до окончательного выяснения генетического словаря [12]. Увеличение содержания $\mathbf{G} + \mathbf{C}$ в ДНК различно отражается на составе аминокислот в белке и характеризуется следующими тремя зависимостями. Содержание одних аминокислот увеличивается при увеличении количества $\mathbf{G} + \mathbf{C}$ в ДНК, содержание других уменьшается, а содержание третьих остается индифферентным относительно количества цитозина и гуанина. В зависимости от содержания гуанина и цитозина дуплеты нуклеотидов, кодирующих аминокислоты разделяются на три класса: не включающие ни гуанин, ни цитозин; содержащие исключительно гуанин и цитозин и промежуточные [12]. Очевидно, что присутствие в кодоне гуанина или цитозина в большой степени детерминирует образование совершенно определенной аминокислоты, в то время как наличие аденина или урацила не является высоко специфичным: например, дуплет $\mathbf{AA}$ соответствует и Lys, и Asn, дуплет $\mathbf{UU}$ кодирует Phen и Leu, в то время как $\mathbf{CG}$ всегда детерминирует Arg).

## 2 Структура матрицы дуплетов

Поскольку основание в третьем положении допускает большую вариабельность для одной и той же аминокислоты, целесообразно рассмотреть первые два основания кодонов $x, y$ отдельно от третьего $z$. При этом шестнадцать возможных дуплетов по способности детерминировать аминокислоту распадаются на два октета. Восемь дуплетов (более “сильных”) однозначно кодирующих аминокислоты независимо от третьего основания, и восемь (“слабых”), у которых третье основание определяет значение кодона. Дуплеты $xy$ первого и второго октетов резко различаются по составу. В первом октете $\mathbf{A}$ встечается лишь один раз, во втором — один раз $\mathbf{C}$ . Причем, переход от дуплетов “сильного” октета в дуплеты “слабого” октета можно получить следующей заменой [17]

$$\mathbf{C} \overset{*}{\Longleftrightarrow} \mathbf{A},\ \mathbf{G} \overset{*}{\Longleftrightarrow} \mathbf{U}, \tag{1}$$

которую мы будем обозначать “операция $(*)$” и называть пурин-пиримидиновой инверсией.

Четыре нуклеотида можно расположить по способности однозначно детерминировать аминокислоты в порядке убывания следующим образом [18]:

$$\begin{array}{cccc} \text{Пиримидин} & \text{Пурин} & \text{Пиримидин} & \text{Пурин} \\ \mathbf{C} & \mathbf{G} & \mathbf{U} & \mathbf{A} \\ \text{очень сильная} & \text{сильная} & \text{слабая} & \text{очень слабая} \end{array} \tag{2}$$

Эти нуклеотиды отличаются и по числу водородных связей, которые они могут образовывать с комплементарными нуклеотидами антикодона: каждая из сильных букв ($\mathbf{C}$ и $\mathbf{G}$) образует по три водородные связи, в то время как каждая из слабых букв ($\mathbf{U}$ и $\mathbf{A}$) образует лишь по две водородные связи [18, 19]. Можно предположить, что, чем больше водородных связей в дуплете, тем меньшее значение имеет взаимодействие $z \to z'$ третьего основания с антикодоном.

Для того, чтобы перейти от качественного описания структуры генетического кода (относительно способности кодировать аминокислоту) к количественному, введем числовую характеристику эмпирической “силы” — “степень детерминации” нуклеотида $d$. Исходя из (2) будем считать, что степень детерминации $d$ может принимать значения от одного до четырех соответственно возрастанию этой силы. Если обозначить степень детерминации кодона верхним индексом в скобках, то четверку оснований (2) можно представить в виде вектора-столбца

$$\mathbb{V} = \begin{pmatrix} \mathbf{C}^{(4)} \\ \mathbf{G}^{(3)} \\ \mathbf{U}^{(2)} \\ \mathbf{A}^{(1)} \end{pmatrix} \tag{3}$$

и соответствующей вектор-строки

$$\mathbb{V}^T = \begin{pmatrix} \mathbf{C}^{(4)} & \mathbf{G}^{(3)} & \mathbf{U}^{(2)} & \mathbf{A}^{(1)} \end{pmatrix}. \tag{4}$$

Операция $(*)$ (1) действует на вектор-столбец $\mathbb{V}$ следующим образом

$$\mathbb{V}^* = \begin{pmatrix} \mathbf{A}^{(1)} \\ \mathbf{U}^{(2)} \\ \mathbf{G}^{(3)} \\ \mathbf{C}^{(4)} \end{pmatrix} \tag{5}$$

Рассмотрим внешнее произведение [20] вектора-столбца (3) и вектора-строки (4)

$$\begin{aligned} \mathbb{M} \ &= \ \mathbb{V} \times \mathbb{V}^T = \begin{pmatrix} \mathbf{C}^{(4)} \\ \mathbf{G}^{(3)} \\ \mathbf{U}^{(2)} \\ \mathbf{A}^{(1)} \end{pmatrix} \begin{pmatrix} \mathbf{C}^{(4)} & \mathbf{G}^{(3)} & \mathbf{U}^{(2)} & \mathbf{A}^{(1)} \end{pmatrix} = \\ &= \begin{pmatrix} \mathbf{C}^{(4)}\mathbf{C}^{(4)} & \mathbf{C}^{(4)}\mathbf{G}^{(3)} & \mathbf{C}^{(4)}\mathbf{U}^{(2)} & \mathbf{C}^{(4)}\mathbf{A}^{(1)} \\ \mathbf{G}^{(3)}\mathbf{C}^{(4)} & \mathbf{G}^{(3)}\mathbf{G}^{(3)} & \mathbf{G}^{(3)}\mathbf{U}^{(2)} & \mathbf{G}^{(3)}\mathbf{A}^{(1)} \\ \mathbf{U}^{(2)}\mathbf{C}^{(4)} & \mathbf{U}^{(2)}\mathbf{G}^{(3)} & \mathbf{U}^{(2)}\mathbf{U}^{(2)} & \mathbf{U}^{(2)}\mathbf{A}^{(1)} \\ \mathbf{A}^{(1)}\mathbf{C}^{(4)} & \mathbf{A}^{(1)}\mathbf{G}^{(3)} & \mathbf{A}^{(1)}\mathbf{U}^{(2)} & \mathbf{A}^{(1)}\mathbf{A}^{(1)} \end{pmatrix}. \end{aligned} \tag{6}$$

Отсюда видно, что $\mathbb{M}$ (6) представляет собой матрицу дуплетов, внутренняя структура которой определяется внешним произведением векторов. Именно этот факт обуславливает высокую степень симметрии матрицы $\mathbb{M}$ и позволяет исследовать свойства математической модели генетического словаря в терминах абстрактной теории групп.

# 3 Свойства матрицы дуплетов и степени детерминации

Будем полагать, что в первом приближении степень детерминации кодона $d$ является аддитивной характеристикой, то есть степень детерминации дуплета складывается из степеней детерминации составляющих его нуклеотидов (например, $d_{\mathbf{GA}} = d_{\mathbf{G}} + d_{\mathbf{A}}$). Тогда из матрицы дуплетов $\mathbb{M}$ можно получить соответствующую (симметрическую) матрицу $D$ степеней детерминации

$$D = \begin{pmatrix} \mathbf{8} & \mathbf{7} & \mathbf{6} & \mathbf{5} \\ \mathbf{7} & \mathbf{6} & \mathbf{5} & \mathbf{4} \\ \mathbf{6} & \mathbf{5} & \mathbf{4} & \mathbf{3} \\ \mathbf{5} & \mathbf{4} & \mathbf{3} & \mathbf{2} \end{pmatrix}, \qquad (7)$$

в которой явно прослеживаются симметрийные закономерности. Отметим, что симметрия матрицы (7) настолько высока, что матрица сингулярна, т. е. ее детерминант равен нулю $\det D = \mathbf{0}$, и ранг матрицы равен двум $\operatorname{rank} D = \mathbf{2}$, ее дефект также равен двум. Это является следствием того факта, что она есть внешнее произведение (6). Примечательно, что след матрицы равен $\operatorname{tr} D = \mathbf{20}$ и совпадает с суммой элементов боковой диагонали. Видно также, что по боковой диагонали матрицы (7) и параллельно ей находятся "равносильные" дуплеты. Операция $(*)$ (1) соответствует отражению матрицы степеней детерминации (7) относительно боковой диагонали

$$D^* = \begin{pmatrix} \mathbf{2} & \mathbf{3} & \mathbf{4} & \mathbf{5} \\ \mathbf{3} & \mathbf{4} & \mathbf{5} & \mathbf{6} \\ \mathbf{4} & \mathbf{5} & \mathbf{6} & \mathbf{7} \\ \mathbf{5} & \mathbf{6} & \mathbf{7} & \mathbf{8} \end{pmatrix}. \qquad (8)$$

и, следовательно, пурин-пиримидиновая инверсия не меняет основных свойств матрицы $D$, то есть имеем $\det D^* = \det D = \mathbf{0}$, $\operatorname{rank} D^* = \operatorname{rank} D = \mathbf{2}$ и $\operatorname{tr} D^* = \operatorname{tr} D = \mathbf{20}.$

Если расположить боковую диагональ матрицы $\mathbb{M}$ горизонтально, то мы получаем ромбическую структуру дуплетов

| | | | | | | | | |
|---|---|---|---|---|---|---|---|---|
| | | | **CC** | | | | $= \mathbf{8}$ | |
| | | **GC** | | **CG** | | | $= \mathbf{7}$ сильные | |
| | **CU** | | **GG** | | **CU** | | $= \mathbf{6}$ | |
| **AC** | | **UG** | | **GU** | | **CA** | $= \mathbf{5}$ переходные | (9) |
| | **AG** | | **UU** | | **GA** | | $= \mathbf{4}$ | |
| | | **AU** | | **UA** | | | $= \mathbf{3}$ слабые | |
| | | | **AA** | | | | $= \mathbf{2}$ | |

соответствующую ромбическому варианту генетического словаря [18, 21], в которой определена "сила" каждого дуплета в терминах его степени детерминации так, что горизонтальные ряды состоят из равносильных дуплетов. Пурин-пиримидиновая инверсия (переход (1)) соответствует отражению ромбической структуры относительно диагонали. Дуплеты со степенью детерминации больше $\mathbf{5}$ кодируют одну аминокислоту, со степенью детерминации меньше $\mathbf{5}$ кодируют две аминокислоты, значение $\mathbf{5}$ — промежуточное: из этого ряда $\mathbf{AC}$ и $\mathbf{GU}$ детерминируют по одной аминокислоте, а $\mathbf{UG}$ и $\mathbf{CA}$ по две. Аминокислоты, кодируемые дуплетами с низкой степенью дерминации (меньше **5**) принадлежат к разным классам по химическому типу радикалов, кроме дуплета $\mathbf{GA}$, детерминирующего Asp и Glu одного химического класса. В промежуточном случае степень детерминации дуплета пропорциональна числу водородных связей. Пуриновые основания ($\mathbf{G}$ и $\mathbf{A}$) образуют по две водородных связи, а пиримидиновые ($\mathbf{C}$ и $\mathbf{U}$) по три связи.

# 4 Трехмерная матрица кодонов

От матрицы дуплетов $\mathbb{M}$ (6) можно перейти к структуре триплетного кода следующим образом. Представим, что матрица дуплетов $\mathbb{M}$ находится в плоскости координат $xy$. Умножим эту матрицу на вектор столбец $\mathbb{V}$ (3), лежащий на оси, перпендикулярной плоскости $xy$, то есть построим тройное внешнее произведение $\mathbb{K} = \mathbb{V} \times \mathbb{M}$. Таким образом, по аналогии с (6) мы получили трехмерную матрицу четвертого порядка, или кубическую матрицу над множеством триплетов.

Для определения степени детерминации триплетов также воспользуемся предположением об аддитивности, как и для дуплетов (например, $d_{\mathbf{CGA}} = d_{\mathbf{C}} + d_{\mathbf{G}} + d_{\mathbf{A}}$). Тогда каждый из 64 элементов (кодонов) кубической матрицы будет иметь числовую характеристику — степень детерминации кодона от **3** до **12**, которую для краткости будем называть силой кодона, а суммы степеней детерминации кодонов, лежащих на одной грани — силой грани. Кодоны с минимальной силой **3** — ($\mathbf{AAA}$) и максимальной **12** — ($\mathbf{CCC}$) лежат в противоположных вершинах куба, причем $\mathbf{CCC}$ находится в плоскости верхней грани, а $\mathbf{AAA}$ — в плоскости нижней грани куба. Назовем $\mathbf{AAA}(\mathbf{3})$-Lys минимальным полюсом, а $\mathbf{CCC}(\mathbf{12})$-Pro— максимальным[1]. Сумма сил по всем четырем диагоналям куба равна **30**, что указывает на высокую степень симметрии матрицы. Сумма сил верхней грани куба

$$\begin{array}{llll} \mathbf{CCC12})-\mathsf{Pro} & \mathbf{CCG}(\mathbf{11})-\mathsf{Pro} & \mathbf{CCU}(\mathbf{10})-\mathsf{Pro} & \mathbf{CCA}(\mathbf{9})-\mathsf{Pro} \\ \mathbf{CGC}(\mathbf{11})-\mathsf{Arg} & \mathbf{CGG}(\mathbf{10})-\mathsf{Arg} & \mathbf{CGU}(\mathbf{9})-\mathsf{Arg} & \mathbf{CGA}(\mathbf{8})-\mathsf{Arg} \\ \mathbf{CUC}(\mathbf{10})-\mathsf{Leu} & \mathbf{CUG}(\mathbf{9})-\mathsf{Leu} & \mathbf{CUU}(\mathbf{8})-\mathsf{Leu} & \mathbf{CUA}(\mathbf{7})-\mathsf{Leu} \\ \mathbf{CAC}(\mathbf{9})-\mathsf{His} & \mathbf{CAG}(\mathbf{8})-\mathsf{Gln} & \mathbf{CAU}(\mathbf{7})-\mathsf{His} & \mathbf{CAA}(\mathbf{6})-\mathsf{Gln} \end{array} \tag{10}$$

равна **144**, а нижней грани

$$\begin{array}{llll} \mathbf{ACC}(\mathbf{9})-\mathsf{Thr} & \mathbf{ACG}(\mathbf{8})-\mathsf{Thr} & \mathbf{ACU}(\mathbf{7})-\mathsf{Thr} & \mathbf{ACA}(\mathbf{6})-\mathsf{Thr} \\ \mathbf{AGC}(\mathbf{8})-\mathsf{Ser} & \mathbf{AGG}(\mathbf{7})-\mathsf{Arg} & \mathbf{AGU}(\mathbf{6})-\mathsf{Ser} & \mathbf{AGA}(\mathbf{5})-\mathsf{Arg} \\ \mathbf{AUC}(\mathbf{7})-\mathsf{Ile} & \mathbf{AUG}(\mathbf{6})-\mathsf{Met} & \mathbf{AUU}(\mathbf{5})-\mathsf{Ile} & \mathbf{AUA}(\mathbf{4})-\mathsf{Ile} \\ \mathbf{AAC}(\mathbf{6})-\mathsf{Asn} & \mathbf{AAG}(\mathbf{5})-\mathsf{Lys} & \mathbf{AAU}(\mathbf{4})-\mathsf{Asn} & \mathbf{AAA}(\mathbf{3})-\mathsf{Lys} \end{array} \tag{11}$$

равна **96**. Боковые грани (наружные плоскости кубической матрицы $\mathbb{K}$) по силе различаются относительно диагонального сечения, построенного через элементы $\mathbf{CAC}(\mathbf{9})$**,** $\mathbf{CCA}(\mathbf{9})$**,** $\mathbf{ACA}(\mathbf{6})$**,** $\mathbf{AAC}(\mathbf{6})$. Грани, имеющие общее ребро $\mathbf{CAA}$— $\mathbf{AAA}$

$$\begin{array}{llll} \mathbf{CAC}(\mathbf{9})-\mathsf{His} & \mathbf{CAG}(\mathbf{8})-\mathsf{Gln} & \mathbf{CAU}(\mathbf{7})-\mathsf{His} & \mathbf{CAA}(\mathbf{6})-\mathsf{Gln} \\ \mathbf{GAC}(\mathbf{8})-\mathsf{Asp} & \mathbf{GAG}(\mathbf{7})-\mathsf{Glu} & \mathbf{GAU}(\mathbf{6})-\mathsf{Asp} & \mathbf{GAA}(\mathbf{5})-\mathsf{Glu} \\ \mathbf{UAC}(\mathbf{7})-\mathsf{Tyr} & \mathbf{UAG}(\mathbf{6})-\mathsf{TERM} & \mathbf{UAU}(\mathbf{5})-\mathsf{Tyr} & \mathbf{UAA}(\mathbf{4})-\mathsf{TERM} \\ \mathbf{AAC}(\mathbf{6})-\mathsf{Asn} & \mathbf{AAG}(\mathbf{5})-\mathsf{Lys} & \mathbf{AAU}(\mathbf{4})-\mathsf{Asn} & \mathbf{AAA}(\mathbf{3})-\mathsf{Lys} \end{array} \tag{12}$$

и

$$\begin{array}{llll} \mathbf{CAA}(\mathbf{6})-\mathsf{Gln} & \mathbf{CUA}(\mathbf{7})-\mathsf{Ley} & \mathbf{CGA}(\mathbf{8})-\mathsf{Arg} & \mathbf{CCA}(\mathbf{9})-\mathsf{Pro} \\ \mathbf{AGC}(\mathbf{5})-\mathsf{Ser} & \mathbf{GUA}(\mathbf{6})-\mathsf{Val} & \mathbf{GGA}(\mathbf{7})-\mathsf{Gly} & \mathbf{GCA}(\mathbf{8})-\mathsf{Ala} \\ \mathbf{UAA}(\mathbf{4})-\mathsf{TERM} & \mathbf{UAA}(\mathbf{5})-\mathsf{Leu} & \mathbf{UGA}(\mathbf{6})-\mathsf{TERM} & \mathbf{UCA}(\mathbf{7})-\mathsf{Ser} \\ \mathbf{AAA}(\mathbf{3})-\mathsf{Lys} & \mathbf{AUA}(\mathbf{4})-\mathsf{Ile} & \mathbf{AGA}(\mathbf{5})-\mathsf{Arg} & \mathbf{ACA}(\mathbf{6})-\mathsf{Thr} \end{array} \tag{13}$$

[1] В скобках приводится сила триплета.

имеют силу, равную **96**, а грани с общим ребром **CCC— ACC**

$$\begin{array}{llll} \mathbf{CCC}(\mathbf{12})-\mathsf{Pro} & \mathbf{CCG}(\mathbf{11})-\mathsf{Pro} & \mathbf{CCU}(\mathbf{10})-\mathsf{Pro} & \mathbf{CCA}(\mathbf{9})-\mathsf{Pro} \\ \mathbf{GCC}(\mathbf{11})-\mathsf{Ala} & \mathbf{GCG}(\mathbf{10})-\mathsf{Ala} & \mathbf{GCU}(\mathbf{9})-\mathsf{Ala} & \mathbf{GCA}(\mathbf{8})-\mathsf{Ala} \\ \mathbf{UCC}(\mathbf{10})-\mathsf{Ser} & \mathbf{UCG}(\mathbf{9})-\mathsf{Ser} & \mathbf{UCU}(\mathbf{8})-\mathsf{Ser} & \mathbf{UCA}(\mathbf{7})-\mathsf{Ser} \\ \mathbf{ACC}(\mathbf{9})-\mathsf{Thr} & \mathbf{ACG}(\mathbf{8})-\mathsf{Thr} & \mathbf{ACU}(\mathbf{7})-\mathsf{Thr} & \mathbf{ACA}(\mathbf{6})-\mathsf{Thr} \end{array} \tag{14}$$

и

$$\begin{array}{llll} \mathbf{CAC}(\mathbf{9})-\mathsf{His} & \mathbf{CUC}(\mathbf{10})-\mathsf{Leu} & \mathbf{CGC}(\mathbf{11})-\mathsf{Arg} & \mathbf{CCC}(\mathbf{12})-\mathsf{Pro} \\ \mathbf{GAC}(\mathbf{8})-\mathsf{Asp} & \mathbf{GUC}(\mathbf{9})-\mathsf{Val} & \mathbf{GGC}(\mathbf{10})-\mathsf{Gly} & \mathbf{GCC}(\mathbf{11})-\mathsf{Ala} \\ \mathbf{UAC}(\mathbf{7})-\mathsf{Tyr} & \mathbf{UUC}(\mathbf{8})-\mathsf{Phe} & \mathbf{UGC}(\mathbf{9})-\mathsf{Cys} & \mathbf{UCC}(\mathbf{10})-\mathsf{Ser} \\ \mathbf{AAC}(\mathbf{6})-\mathsf{Asn} & \mathbf{AUC}(\mathbf{7})-\mathsf{Ile} & \mathbf{AGC}(\mathbf{8})-\mathsf{Ser} & \mathbf{ACC}(\mathbf{9}-\mathsf{Thr} \end{array} \tag{15}$$

имеют силу, равную **144**. Следовательно, грани имеющие общую точку $\mathbf{AAA}(\mathbf{3})$ характеризуются силой **96**, а грани, пересекающиеся в точке максимального полюса $\mathbf{CCC}(\mathbf{12})$, имеют силу **144** каждая. Это подтверждает высокую симметрию кубической матрицы $\mathbb{K}$.

Внутренние плоскости матрицы, параллельные граням, число которых шесть, как и наружных, по сумме сил разделяются на две группы: с силами **112** и **128**. Две вертикальные плоскости $[\mathbf{CAU}(\mathbf{7}), \mathbf{CCU}(10), \mathbf{ACU}(\mathbf{7}), \mathbf{AAU}(4)]$ и $[\mathbf{CUC}(10), \mathbf{CUA}(\mathbf{7}), \mathbf{AUA}(4), \mathbf{AUC}(\mathbf{7})]$, линия пересечения которых проходит через $\mathbf{CUU}(8)$— $\mathbf{AUU}(\mathbf{5})$, а также плоскость $[\mathbf{UAC}(\mathbf{7}), \mathbf{UCC}(10), \mathbf{UCA}(\mathbf{7}), \mathbf{UAA}(4)]$, пересекающая две первые и имеющая с ними общую точку $\mathbf{UUU}(\mathbf{6})$ имеют силу **112** каждая и находятся ближе к минимальному полюсу.

Плоскости $[\mathbf{CGC}(11), \mathbf{CGA}(8), \mathbf{AGA}(5), \mathbf{AGC}(8)]$ и $[\mathbf{CAC}(8), \mathbf{CCG}(11), \mathbf{ACG}(8), \mathbf{AAC}(5)]$ пересекаются по следующей прямой $\mathbf{CGG}(10)$— $\mathbf{AGG}(\mathbf{7})$ и имеют с плоскостью $[\mathbf{GAC}(8), \mathbf{GCC}(11), \mathbf{GCA}(8), \mathbf{GAA}(\mathbf{5})]$ общую точку $\mathbf{GGG}(\mathbf{9})$. Сила каждой из них равна **128**.

Отметим что, точками пересечения равносильных плоскостей являются монотонные тринуклеотиды, находящиеся на главной диагонали. Отсюда следует правило:

> *Сила грани или плоскости, параллельной грани,* $d_{plane}$ *однозначно определяется ее единственным монотонным тринуклеотидом* $\mathbf{AAA}(\mathbf{3})$, $\mathbf{UUU}(\mathbf{6})$, $\mathbf{GGG}(\mathbf{9})$, $\mathbf{CCC}(\mathbf{12})$ *и равна* **96**, **112**, **128**. **144** *соответственно.*

Важно, что график зависимости силы плоскости $d_{plane}$ от силы принадлежащего ей монотонного тринуклеотида $d_{mono}$ представляет собой прямую линию

$$d_{plane} = \frac{16}{3} d_{mono} + 80, \tag{16}$$

что является следствием высокой симметрии кубической матрицы.

# 5 Кубическая матрица кодонов и вырожденность кода

Рассмотрим особенности расположения кодонов и их значений в построенной кубической матрице $\mathbb{K}$. Поскольку каждому кодону, кроме трех терминальных, соответствует аминокислота, то можно говорить о трехмерной модели генетического словаря.

Отметим следующие свойства этой модели. Гидрофобные аминокислоты Leu, Val, Phe, Ile лежат в одной плоскости $\mathbf{CUC}(10), \mathbf{CUA}(\mathbf{7}), \mathbf{AUA}(4), \mathbf{AUC}(\mathbf{7})$. Оксимоноаминокарбоновые кислоты серин (Ser) и треонин (Thr) находятся на наружной грани $\mathbf{CCC}(\mathbf{12}), \mathbf{CCA}(\mathbf{9}), \mathbf{ACA}(\mathbf{6}), \mathbf{ACC}(\mathbf{9})$. На верхней грани куба находятся четыре из шести кодонов лейцина и аргинина, и все кодоны этой грани содержат цитозин. Кодоны пролина занимают строку с наибольшими степенями детерминации $\mathbf{CCC}(\mathbf{12}), \mathbf{CCG}(\mathbf{11}), \mathbf{CCU}(\mathbf{10}), \mathbf{CCA}(\mathbf{9})$. Аминокислоты, кодирующиеся четырьмя кодонами, расположены в строку, причем моноаминокарбоновые аминокислоты лежат в одной плоскости

$$\begin{array}{llll} \mathbf{GCC(11)}-\mathsf{Ala} & \mathbf{GCG(10)}-\mathsf{Ala} & \mathbf{GCU(9)}-\mathsf{Ala} & \mathbf{GCA(8)}-\mathsf{Ala} \\ \mathbf{GGC(10)}-\mathsf{Gly} & \mathbf{GGG(9)}-\mathsf{Gly} & \mathbf{GGU(8)}-\mathsf{Gly} & \mathbf{GGA(7)}-\mathsf{Gly} \\ \mathbf{GUC(9)}-\mathsf{Val} & \mathbf{GUG(8)}-\mathsf{Val} & \mathbf{GUU(7)}-\mathsf{Val} & \mathbf{GUA(6)}-\mathsf{Val} \\ \mathbf{GAC(8)}-\mathsf{Asp} & \mathbf{GAG(7)}-\mathsf{Glu} & \mathbf{GAU(6)}-\mathsf{Asp} & \mathbf{GAA(5)}-\mathsf{Glu} \end{array} \qquad (17)$$

Последнюю строку занимают моноаминодикарбоновые аминокислоты: аспарагиновая и глутаминовая, обладающие гидрофильными свойствами. Уникальные кодоны $\mathbf{AUG(6)}-\mathsf{Met}$ находится в плоскости нижнего основания основания матрицы, а $\mathbf{UGG(8)}-\mathsf{Trp}$ лежит в одной плоскости

$$\begin{array}{llll} \mathbf{UCC(10)}-\mathsf{Ser} & \mathbf{UCG(9)}-\mathsf{Ser} & \mathbf{UCU(8)}-\mathsf{Ser} & \mathbf{UCA(7)}-\mathsf{Ser} \\ \mathbf{UGC(9)}-\mathsf{Cys} & \mathbf{UGG(8)}-\mathsf{Trp} & \mathbf{UGU(7)}-\mathsf{Cys} & \mathbf{UGA(6)}-\mathsf{TERM} \\ \mathbf{UUC(8)}-\mathsf{Phe} & \mathbf{UUG(7)}-\mathsf{Leu} & \mathbf{UUU(6)}-\mathsf{Phe} & \mathbf{UUA(5)}-\mathsf{Leu} \\ \mathbf{UAC(7)}-\mathsf{Tyr} & \mathbf{UAG(6)}-\mathsf{TERM} & \mathbf{UAU(5)}-\mathsf{Tyr} & \mathbf{UAA(4)}-\mathsf{TERM} \end{array} \qquad (18)$$

с терминальными кодонами $\mathbf{UAA(4), UGA(6), UAG(6)}$, которые имеют степени детерминации не больше **6** и находятся на равносильных боковых гранях с $d = \mathbf{96}$. В общем случае можно заметить, что, чем больше кодонов кодирует одну аминокислоту, тем больше сила каждого из них (**8-10**), уникальные кодоны, напротив, имеют невысокие степени детерминации (**4-8**).

Определим силу аминокислоты $d_{\mathsf{AMK}}$ как среднее арифметическое сил кодонов $d_{codon}$, детерминирующих ее

$$d_{\mathsf{AMK}} = \frac{\sum d_{codon}}{n_{deg}}, \qquad (19)$$

где $n_{deg}$ — это ее степень вырожденности. То есть, для каждой из 20 аминокислот мы получили числовую характеристику $d_{\mathsf{AMK}}$, которая показывает с какой силой детерминируется данная аминокислота. Эта зависимость и зависимость средней силы аминокислоты $d_{\mathsf{AMK}}$ от числа кодонов $n_{deg}$ представлены в Таблице 1.

| AMK | $d_{\mathsf{AMK}}$ | $n_{deg}$ |
|---|---|---|
| Lys | $\mathbf{4}$ | 2 |
| Asn | $\mathbf{5}$ | 2 |
| Ile | $\mathbf{5\frac{1}{3}}$ | 3 |
| Glu/Met/Tyr | $\mathbf{6}$ | 2/1/2 |
| Phe/Asp/Gln | $\mathbf{7}$ | 2/2/2 |
| Val/Thr | $\mathbf{7\frac{1}{2}}$ | 4/4 |
| Leu | $\mathbf{7\frac{2}{3}}$ | 6 |
| Cys/Trp/Ser/His | $\mathbf{8}$ | 2/1/6/2 |
| Arg | $\mathbf{8\frac{1}{3}}$ | 6 |
| Gly | $\mathbf{8\frac{1}{2}}$ | 4 |
| Ala | $\mathbf{9\frac{1}{2}}$ | 4 |
| Pro | $\mathbf{11}$ | 4 |

Это позволяет анализировать различные свойства аминокислот в зависимости от введенной силы (степени детерминируемости $d_{AMK}$).

# 6 Свойства равносильных сечений

Рассмотрим геометрическое расположение в кубической матрице $\mathbb{K}$ кодонов, имеющих равную силу. Кодоны с одинаковой силой $d_{codon} = const$ лежат в плоскостях, перпендикулярных главной диагонали куба $\mathbf{AAA(3)}$— $\mathbf{CCC(12)}$. Можно показать, что таких плоскостей (или сечений) десять (с учетом двух угловых одноэлементных), которые представляются в виде различных геометрических фигур. Сила сечения $d_{section}$ определяется как сила каждого из входящих в него кодонов $d_{section} = d_{codon} = const$, а количество элементов в сечении находятся в следующей зависимости от $d_{section}$

| Сила сечения $d_{section}$ | **3** | **4** | **5** | **6** | **7** | **8** | **9** | **10** | **11** | **12** |
|---|---|---|---|---|---|---|---|---|---|---|
| Число элементов в сечении | 1 | 3 | 6 | 10 | 12 | 12 | 10 | 6 | 3 | 1 |

Отсюда видно, что десять сечений распадаются на пять взаимодополнительных пар, имеющих одинаковую форму и количество триплетов и связанных между собой пурин-пиримидиновой инверсией и отражением. Если силу сечения дополнительной пары обозначить $d^*_{section}$, то из Таблицы 2 следует формула

$$d_{section} + d^*_{section} = \mathbf{15}. \tag{20}$$

В простейшем случае одноэлементной пары наблюдается монотонная пурин-пиримидиновая инверсия $\mathbf{AAA(3)} \overset{*}{\Longleftrightarrow} \mathbf{CCC(12)}$. Следующая пара взаимодополнительных (в смысле формулы (20)) сечений состоит из плоскостей с тремя кодонами

$\underline{d = \mathbf{4}}$ $\qquad$ $\underline{d = \mathbf{11}}$

**UAA** Term $\quad \overset{*}{\Leftrightarrow} \quad$ **CGC** Arg $\quad$ **CCG** Pro

**AAU** Asn $\quad$ **AUA** Ile $\quad$ **GCC** Ala

Отметим, что сечение $d = \mathbf{4}$ не содержит “сильных” дуплетов, а сечение $d = \mathbf{11}$ не содержит “слабых” дуплетов. В остальных парах происходит смешение “слабых” и “сильных” дуплетов. Например, пара сечений с 6 элементами имеет вид

$\underline{d = \mathbf{5}}$

**GAA** Glu

**UAU** Tyr $\quad$ **UUA** Leu

**AAG** Lys $\quad$ **AUU** Ile $\quad$ **AGA** Arg

$\Updownarrow *$

$\underline{d = \mathbf{10}}$

**CUC** Leu $\quad$ **CGG** Arg $\quad$ **CCU** Pro

**GGC** Gly $\quad$ **GCG** Ala

**UCC** Ser

Видно, что сечение $d = \mathbf{4}$ ($d = \mathbf{5}$) является отображением в сечение $d = \mathbf{11}$ ($d = \mathbf{10}$) при действии пурин-пиримидиновой инверсии $\mathbf{G} \overset{*}{\Longleftrightarrow} \mathbf{U}, \mathbf{C} \overset{*}{\Longleftrightarrow} \mathbf{A}$ (1).

Примечательно, что подобные отображения и зеркальная симметрия с поворотом на угол $\pi$ относительно оси $y$ наблюдаются во всех парах геометрически подобных сечений с одинаковым числом элементов, что наглядно прослеживается из

$$\begin{array}{c} d = \mathbf{5}: \ \mathbf{AAG} \to \mathbf{UAU} \to \mathbf{GAA} \to \mathbf{UUA} \to \mathbf{AGA} \to \mathbf{AUU} \\ \Updownarrow * \\ d = \mathbf{10}: \ \mathbf{CCU} \to \mathbf{GCG} \to \mathbf{UCC} \to \mathbf{GGC} \to \mathbf{CUC} \to \mathbf{CGG} \end{array} \tag{21}$$

Пара сечений с силами $\mathbf{6}$ и $\mathbf{9}$ содержит в том числе и монотонные кодоны $\mathbf{UUU}$ и $\mathbf{GGG}$, находящиеся на диагонали куба и внутри сечения

$$\begin{array}{c} d = \mathbf{6}: \ \mathbf{UUU} \to \mathbf{AAC} \to \mathbf{UAG} \to \mathbf{GAU} \to \mathbf{CAA} \to \mathbf{GUA} \\ \to \mathbf{UGA} \to \mathbf{ACA} \to \mathbf{AGU} \to \mathbf{AUG} \\ \Updownarrow * \\ d = \mathbf{9}: \ \mathbf{GGG} \to \mathbf{CCA} \to \mathbf{GCU} \to \mathbf{UCG} \to \mathbf{ACC} \to \mathbf{UGC} \\ \to \mathbf{GUC} \to \mathbf{CAC} \to \mathbf{CUG} \to \mathbf{CGU} \end{array}$$

Сечения с $d = \mathbf{7}$ и $d = \mathbf{8}$ содержат по три элемента, лежащих внутри периметра сечения

$$\begin{array}{c} d = \mathbf{7}: \ \mathbf{AUC} \to \mathbf{UAC} \to \mathbf{GAG} \to \mathbf{CAU} \to \mathbf{CUA} \to \mathbf{GGA} \\ \to \mathbf{UCA} \to \mathbf{ACU} \to \mathbf{AGG} \to \mathbf{UUG} \to \mathbf{GUU} \to \mathbf{UGU} \\ \Updownarrow * \\ d = \mathbf{8}: \ \mathbf{CGA} \to \mathbf{GCA} \to \mathbf{UCU} \to \mathbf{ACG} \to \mathbf{AGC} \to \mathbf{UUC} \\ \to \mathbf{GAC} \to \mathbf{CAG} \to \mathbf{CUU} \to \mathbf{GGU} \to \mathbf{UGG} \to \mathbf{GUG} \end{array}$$

## 7 Геометрические симметрии кубической матрицы кодонов

Рассмотрим сечение с силой $d = \mathbf{4}$, включающее три элемента. Представим его в виде равностороннего треугольника в вершинах которого расположены кодоны (перечислены по часовой стрелке, начиная с нижней левой вершины):

$$\mathbf{AAU} \to \mathbf{UAA} \to \mathbf{AUA}. \tag{22}$$

Элементы сечения имеют связность, так как кодоны отличаются друг от друга одним основанием.

Далее, каждый кодон можно в свою очередь представить как равносторонний треугольник, в вершинах которого лежат единичные нуклеотиды. Вращением, например, треугольника $\mathbf{AAU}$ вокруг оси $z$, препендикулярной его плоскости на углы $\frac{2}{3}\pi$ и $\frac{4}{3}\pi$ можно получить последовательные превращения (переходы) элементов сечения (22).

Поворотом на положительный угол, относительно направленной оси, мы, как это общепринято [22], будем считать поворот, соответствующий вращению правого винта, то есть по часовой стрелке, если смотреть вдоль оси в положительном направлении. Обозначим операцию вращения треугольника $\mathbf{AAU}$ на угол $\frac{2}{3}\pi$ как $R_1$, а $R_2$ — поворот на $\frac{4}{3}\pi$, поворот $\mathbf{AAU}$ на нулевой угол обозначим $E$. Можно показать, что совокупность операций вращения $E, R_1, R_2$ вокруг оси $z$ образует группу. Для доказательства этого утверждения составим таблицу умножения (таблицу Кэли). Согласно определению [22] совокупность $\Gamma$ элементов $G_1, G_2, G_3...$ называется группой, если задан закон "умножения" или операция, удовлетворяющая следующим требованиям. Результат умножения двух элементов $G_a$ и $G_b$ называется произведением и должен принадлежать группе. В нашем случае элементами группы являются повороты $E, R_1, R_2$, а произведением — последовательное выполнение поворотов. Если поворот $R_1$ переводит систему из положения А в положение В, а поворот $R_2$ из положения В в положение С, то произведение $R_1 R_2$ переводит систему из А в С (закон умножения операций вращения). При повороте на нулевой угол треугольник переходит сам в себя, то есть $E$ — операция тождественная, полностью соответствующая требованиям единичного элемента

$$R_1 \times E = R_1\ ;\ R_2 \times E = R_2.$$

Двойное выполнение поворота на $\frac{2}{3}\pi$ идентично повороту на $\frac{4}{3}\pi$, поэтому $R_1 \times R_1 = R_2$. Аналогично находим произведения остальных элементов.

| $G_a \setminus G_b$ | $E$ | $R_1$ | $R_2$ |
|---|---|---|---|
| $E$ | $E$ | $R_1$ | $R_2$ |
| $R_1$ | $R_1$ | $R_2$ | $R_1$ |
| $R_2$ | $R_2$ | $E$ | $R_1$ |

Видно, что совокупность эементов Г является группой, поскольку выполняются условия:

1. Произведение любых двух элементов принадлежит совокупности.

2. Условие ассоциативности, то есть при перемножении трех элементов не должно иметь значения в какой последовательности выполняется это умножение. Поскольку элемент группы — поворот, то это условие выполняется.

3. Единичный элемент оговорен выше, а существование обратного вытекает из таблицы умножения. Как произведение двух операций совмещения, так и действие обратное любой из них, естественно, являются операциями совмещения.

Таким образом мы доказали, что сечение со степенью детерминации $d = \mathbf{4}$ обладает свойствами группы. Аналогично можно доказать групповые свойства сечения $d = \mathbf{11}$, также содержащее три кодона

$$\mathbf{CCG} \to \mathbf{CGC} \to \mathbf{GCC} \tag{23}$$

В каждом из сечений можно найти общее число пуриновых и пиримидиновых оснований.

| $d_{section}$ | $n_{\mathbf{C}}$ | $n_{\mathbf{G}}$ | $n_{\mathbf{U}}$ | $n_{\mathbf{A}}$ | $n_{\mathbf{C}} + n_{\mathbf{U}}$ | $n_{\mathbf{G}} + n_{\mathbf{A}}$ | $\Delta n$ |
|---|---|---|---|---|---|---|---|
| **3** | 0 | 0 | 0 | 3 | 0 | 3 | -3 |
| **4** | 0 | 0 | 3 | 6 | 3 | 6 | -3 |
| **5** | 0 | 3 | 6 | 9 | 6 | 12 | -6 |
| **6** | 3 | 6 | 9 | 12 | 12 | 18 | -6 |
| **7** | 6 | 9 | 12 | 9 | 18 | 18 | 0 |
| **8** | 9 | 12 | 9 | 6 | 18 | 18 | 0 |
| **9** | 12 | 9 | 6 | 3 | 18 | 12 | 6 |
| **10** | 9 | 6 | 3 | 0 | 12 | 6 | 6 |
| **11** | 6 | 3 | 0 | 0 | 6 | 3 | 3 |
| **12** | 3 | 0 | 0 | 0 | 3 | 0 | 3 |

График зависимости $\Delta n$ от $d$ представляет собой периодическую функцию с областью определения {**3**;**12**} и областью значений на отрезке {**-6**;**6**}. Эта функция описывает колебания разности количества пиримидинов и пуринов в кодонах с одинаковыми степенями детерминации, что позволяет говорить об определенном биологическом смысле введенной нами степени детерминации нуклеотида $d$.

Отметим, что речь идет о зависимостях внутри генетического словаря, применимых и к одноцепочечным нуклеотидным последовательностям в отличие от Чаргаффа[23], описавшего пурин-пиримидиновые соотношения в молекуле ДНК. Таким образом, благодаря введенной степени детерминации нуклеотида возможно анализировать не только относительное и качественное, но и абсолютное содержание определенных нуклеотидов в любом генетическом тексте.

# 8 Выводы

Таким образом, алгебраический подход дает возможность по новому взглянуть на проблему генетического кода. Симметрии, наблюдаемые в коде проявляют себя в процессах выбора кодонов для определения различных аминокислот. Предложенная модель построена на основе различной способности нуклеотидов $\mathbf{C}, \mathbf{G}, \mathbf{U}, \mathbf{A}$ однозначно детерминировать аминокислоту. Введенное понятие степени детерминации нуклеотида позволяет представлять генетические тексты как последовательность чисел от 1 до 4 (фактически в четверичной системе счисления). Анализ таких последовательностей может привести к более глубокому пониманию процессов транскрипции и, возможно, к формулировке новых принципов конструирования рекомбинантных ДНК, что в настоящее время является неотъемлимым компонентом развития методов клонирования и генной инженерии.

# Список литературы